\pdfoutput=1
\documentclass[aps,prl,floatfix,superscriptaddress,twocolumn]{revtex4-1}

\usepackage{macros_BW}
\newcommand{\bilayer}{[(CuF$_2$(H$_2$O)$_2$)$_2$pyz]\xspace}
\newcommand{\singlelayer}{[CuF$_2$(H$_2$O)$_2$pyz]\xspace}

\begin{document}

\title{Giant pressure dependence and dimensionality switching in a 
metal-organic quantum antiferromagnet}

\author{B. Wehinger}
\email[]{bjorn.wehinger@unige.ch}
\affiliation{Department of Quantum Matter Physics, University of Geneva, 24, 
Quai Ernest Ansermet, CH-1211 Gen\`eve, Switzerland}
\affiliation{Laboratory for Neutron Scattering and Imaging, Paul Scherrer 
Institute, CH-5232 Villigen-PSI, Switzerland}

\author{C. Fiolka}
\affiliation{Department of Chemistry and Biochemistry, University of Bern, 
Freiestrasse 3, CH-3012 Bern, Switzerland}

\author{A. Lanza}
\affiliation{Department of Chemistry and Biochemistry, University of Bern, 
Freiestrasse 3, CH-3012 Bern, Switzerland}

\author{R. Scatena}
\affiliation{Department of Chemistry and Biochemistry, University of Bern, 
Freiestrasse 3, CH-3012 Bern, Switzerland}

\author{M. Kubus}
\affiliation{Department of Chemistry and Biochemistry, University of Bern, 
Freiestrasse 3, CH-3012 Bern, Switzerland}

\author{A. Grockowiak}
\affiliation{National High Magnetic Field Laboratory, 1800 E. Paul Dirac 
Drive, Tallahassee, FL 32310, USA}

\author{W. A. Coniglio}
\affiliation{National High Magnetic Field Laboratory, 1800 E. Paul Dirac 
Drive, Tallahassee, FL 32310, USA}

\author{D. Graf}
\affiliation{National High Magnetic Field Laboratory, 1800 E. Paul Dirac 
Drive, Tallahassee, FL 32310, USA}

\author{M. Skoulatos}
\affiliation{Heinz-Maier-Leibnitz Zentrum and Physics Department, Technische 
Universit\"at M\"unchen, Lichtenbergstr.~1, 85748 Garching, Germany}

\author{J.-H. Chen}
\affiliation{Condensed Matter Theory Group, Paul Scherrer Institute, CH-5232 
Villigen-PSI, Switzerland}
\affiliation{Theoretical Physics, ETH Z\"urich, CH-8093 Z\"urich, Switzerland}

\author{J. Gukelberger}
\affiliation{Theoretical Physics, ETH Z\"urich, CH-8093 Z\"urich, Switzerland}
\affiliation{D\'epartement de Physique and Institut Quantique, Universit\'e de 
Sherbrooke, Sherbrooke, Qu\'ebec, J1K 2R1, Canada}

\author{N. Casati}
\affiliation{Swiss Light Source, Paul Scherrer Institute, CH-5232 Villigen-PSI, 
Switzerland}

\author{O. Zaharko}
\affiliation{Laboratory for Neutron Scattering and Imaging, Paul Scherrer 
Institute, CH-5232 Villigen-PSI, Switzerland}

\author{P. Macchi}
\affiliation{Department of Chemistry and Biochemistry, University of Bern, 
Freiestrasse 3, CH-3012 Bern, Switzerland}

\author{K. W. Kr\"amer}
\affiliation{Department of Chemistry and Biochemistry, University of Bern, 
Freiestrasse 3, CH-3012 Bern, Switzerland}

\author{S. Tozer}
\affiliation{National High Magnetic Field Laboratory, 1800 E. Paul Dirac 
Drive, Tallahassee, FL 32310, USA}

\author{C. Mudry}
\affiliation{Condensed Matter Theory Group, Paul Scherrer Institute, CH-5232 
Villigen-PSI, Switzerland}

\author{B. Normand}
\affiliation{Division Research with Neutrons and Muons, Paul Scherrer Institute, 
CH-5232 Villigen-PSI, Switzerland}

\author{Ch. R\"uegg}
\affiliation{Department of Quantum Matter Physics, University of Geneva, 24, 
Quai Ernest Ansermet, CH-1211 Gen\`eve, Switzerland}
\affiliation{Division Research with Neutrons and Muons, Paul Scherrer Institute, 
CH-5232 Villigen-PSI, Switzerland}

\date{\today}

\begin{abstract}
We report an extraordinary pressure dependence of the magnetic interactions in 
the metal-organic system [(CuF$_2$(H$_2$O)$_2$)$_2$pyrazine]. At zero pressure, 
this material realizes a quasi-two-dimensional (Q2D) spin-1/2 square-lattice 
Heisenberg antiferromagnet. By high-pressure, high-field susceptibility 
measurements we show that the dominant exchange parameter is reduced 
continuously by a factor of 2 upon compression. Above 18\,kbar, a 
phase transition occurs, inducing an orbital re-ordering that switches the 
dimensionality, transforming the Q2D lattice into weakly coupled chains (Q1D). 
We explain the microscopic mechanisms for both phenomena by combining detailed 
x-ray and neutron diffraction results with quantitative modeling using spin-polarized 
density functional theory. 
\end{abstract}

\maketitle

Quantum fluctuations are especially strong in low-dimensional 
systems, giving rise to numerous exotic phenomena in quantum magnetism
\cite{dagotto_science_1996,Thielemann2009,Savary2017}. The design and 
control of materials with quasi-one- (Q1D) and quasi-two-dimensional 
(Q2D) antiferromagnetic (AFM) interactions is of particular interest for 
potential applications in AFM spintronics, where energy efficiencies are 
outstanding compared to ferromagnets and the spin dynamics is faster by orders
of magnitude \cite{jungwirth_nnano_2016,kosub_nc_2017,wadley_science_2016}. 
A full exploitation of this potential requires further progress in theoretical, 
experimental, and materials physics, specifically designer low-dimensional 
materials with experimentally controlled magnetic exchange to benchmark 
accurate theoretical descriptions.

Metal-organic compounds based on Cu$^{2+}$ ions make excellent model quantum 
magnets because of their localized spin-1/2 moments and large charge gap. 
Suitable materials are based on coordination polymers with rigid linkers 
such as pyrazine (pyz), which provide Cu$^{2+}$ networks with exchange 
parameters on the scale of 0.1-10 K that are robust and strongly anisotropic 
in space \cite{conner_ltp_2006,goddard_njp_2008}. These interactions can be 
determined to high accuracy from thermodynamic and spectroscopic measurements, 
and interaction control can be achieved by chemistry or physics. Chemical 
variation of ligands and counter-ions allows for significant modification 
\cite{woodward_ic_2007,lancaster_prl_2014}, to the point of dimensionality 
control \cite{goddard_prl_2012}, while fine-tuning is possible by isotopic 
substitution \cite{gobbard_prb_2008}. Physically, an applied pressure 
provides direct control of structural and, in turn, magnetic properties 
\cite{musfeldt_ic_2011,lanza_cc_2014}. 

In this Letter, we report on the behavior of \bilayer under pressure. 
Magnetic susceptibility measurements show a massive and continuous change 
of the dominant exchange parameter in two different low-dimensional 
realizations, a Q2D spin-1/2 square-lattice antiferromagnet at pressures up 
to 18\,kbar and Q1D AFM chains at higher pressures. This giant pressure 
dependence is extreme compared to all reported organic and inorganic materials 
\cite{merchant_np_2014}. By diffraction studies and quantitative modeling 
using spin-polarized density functional theory (DFT), we show that its 
origin lies in the pressure-sensitivity of superexchange paths involving 
water ligands. Our results allow unprecedented control of magnetic 
interactions and thus represent an important step towards materials 
choices for quantum magnetism by design. 

Single crystals of \bilayer were grown as described in Sec.~S1 of the 
Supplemental Material (SM) \cite{sm}. Magnetic susceptibility measurements 
were performed using a Tunnel Diode Oscillator (TDO), as detailed in Sec.~S2 
of the SM \cite{sm}, while the magnetic exchange was controlled by isotropic 
compression of a sample aligned with the crystallographic $a$-axis parallel 
to the field. We performed two independent experiments using (i) a piston 
cylinder cell for pressures up to 17.9\,kbar in fields up to 35\,T and 
temperatures down to 1.5\,K and (ii) a specially designed Moissanite anvil 
cell for pressures up to 37.1\,kbar with maximum field 18\,T and minimum 
temperature 0.4\,K.

\begin{figure}[t]
\includegraphics[width=8.5cm]{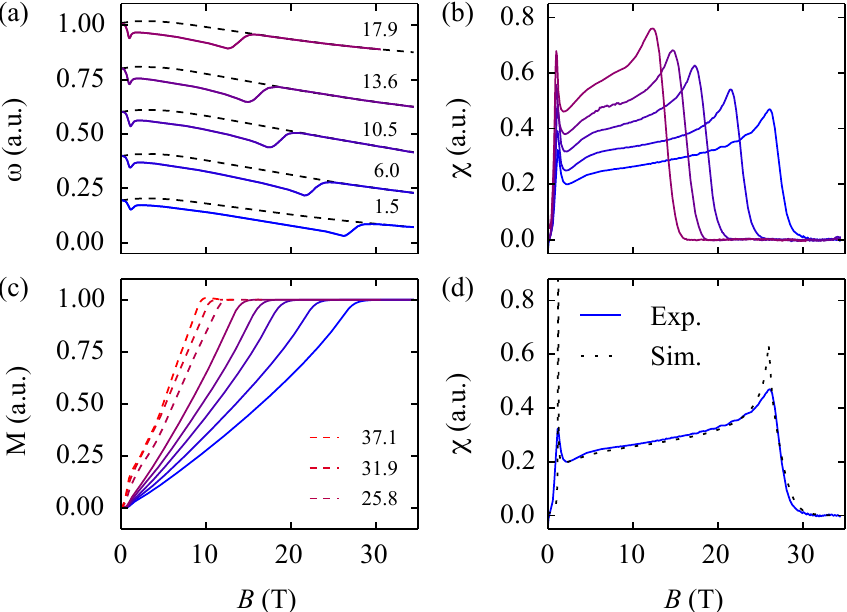}
\caption{(a) Measured TDO resonance frequencies at selected pressures (kbar) for 
$T = 1.5$\,K (solid lines), shown with the magnetoresistive background of 
the resonator coil (dashed). (b) Magnetic susceptibility for the same $T$ 
and $P$ values. (c) Magnetization at 37.1, 31.9, and 25.8\,kbar (dashed 
lines), measured at 0.4\,K, and at the pressures shown in panels (a) and 
(b) (full lines), measured at 1.5\,K. (d) Magnetic susceptibility at 
1.5\,kbar and 1.5\,K as obtained from experiment (full line) and from QMC 
simulations for a system of $32 \times 32 \times 32$ spins (dashed line).} 
\label{fig:bilayer_data}
\end{figure}

The TDO resonance frequency is shown in Fig.~\ref{fig:bilayer_data}(a) as 
a function of field at five different pressures and a constant temperature 
of 1.5\,K. The magnetic susceptibility, $\chi = \partial M/\partial H$ in 
Fig.~\ref{fig:bilayer_data}(b), was obtained by subtracting the 
magnetoresistive background of the resonator coil from the resonance 
frequency. The peak observed at low fields is due to a spin-flop transition, 
occuring at $B_{\rm sf} = 1.2$\,T at 1.5\,kbar and shifting to 1.0\,T at 
17.9\,kbar. Otherwise $\chi$ shows a gradual increase with field and a 
pronounced peak prior to saturation. The magnetization 
[Fig.~\ref{fig:bilayer_data}(c)], obtained by integrating $\chi$, changes 
little for fields below $B_{\rm sf}$, then shows increasing field-alignment 
up to a saturation field $B_c$ that changes dramatically with pressure.

\begin{figure}[t]
\includegraphics[width=8.5cm]{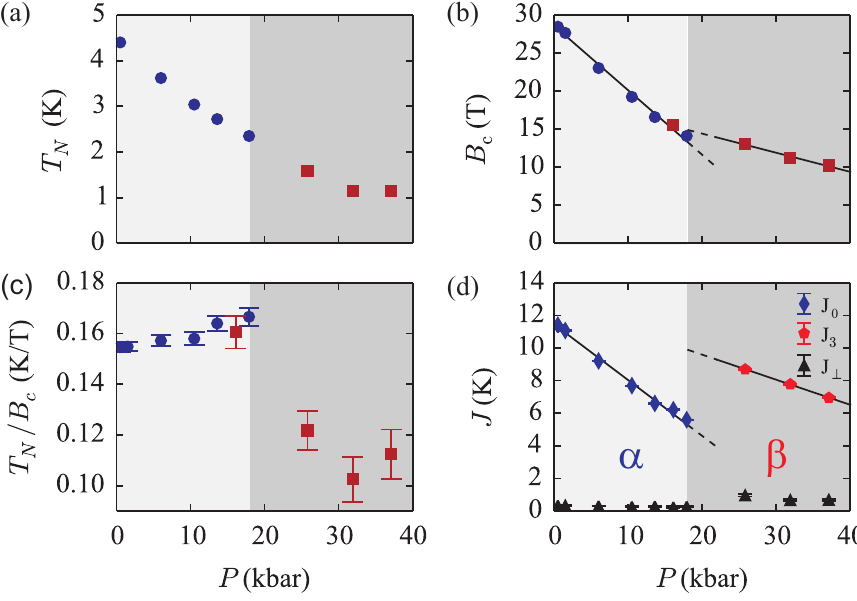}
\caption{(a) N\'eel temperature $T_N$ measured as a function of pressure $P$. 
(b) Saturation field $B_c$ obtained from self-consistent fitting procedure; 
black lines show linear fits. Dark blue circles show data obtained using the 
piston cell, dark red squares using the Moissanite cell. (c) Ratio $T_N/B_c$ as 
a function of $P$, illustrating the evolution of dimensionality; the sharp 
drop marks the phase transition to the Q1D magnetic system. (d) Exchange parameters
obtained from QMC fits to the experimental data together with linear fit
(black lines).}
\label{fig:bilayer_exchange}
\end{figure}

The N\'eel temperature, $T_N$ in Fig.~\ref{fig:bilayer_exchange}(a), was 
determined by measuring the temperature dependence of the resonance frequency 
at $B_{\rm sf}$, which allows for a precise measurement of $T_N$ because the 
changes are particularly pronounced at resonance. The relative change of 
$T_N$ with pressure is also dramatic, and quite unprecedented over such a 
pressure range. We note that $T_N$ is a significant fraction of our 
measurement temperature, and thus care is required to extract the underlying 
magnetic exchange parameters from a consistent fitting procedure. 

\begin{figure}[t]
\includegraphics[width=8.5cm]{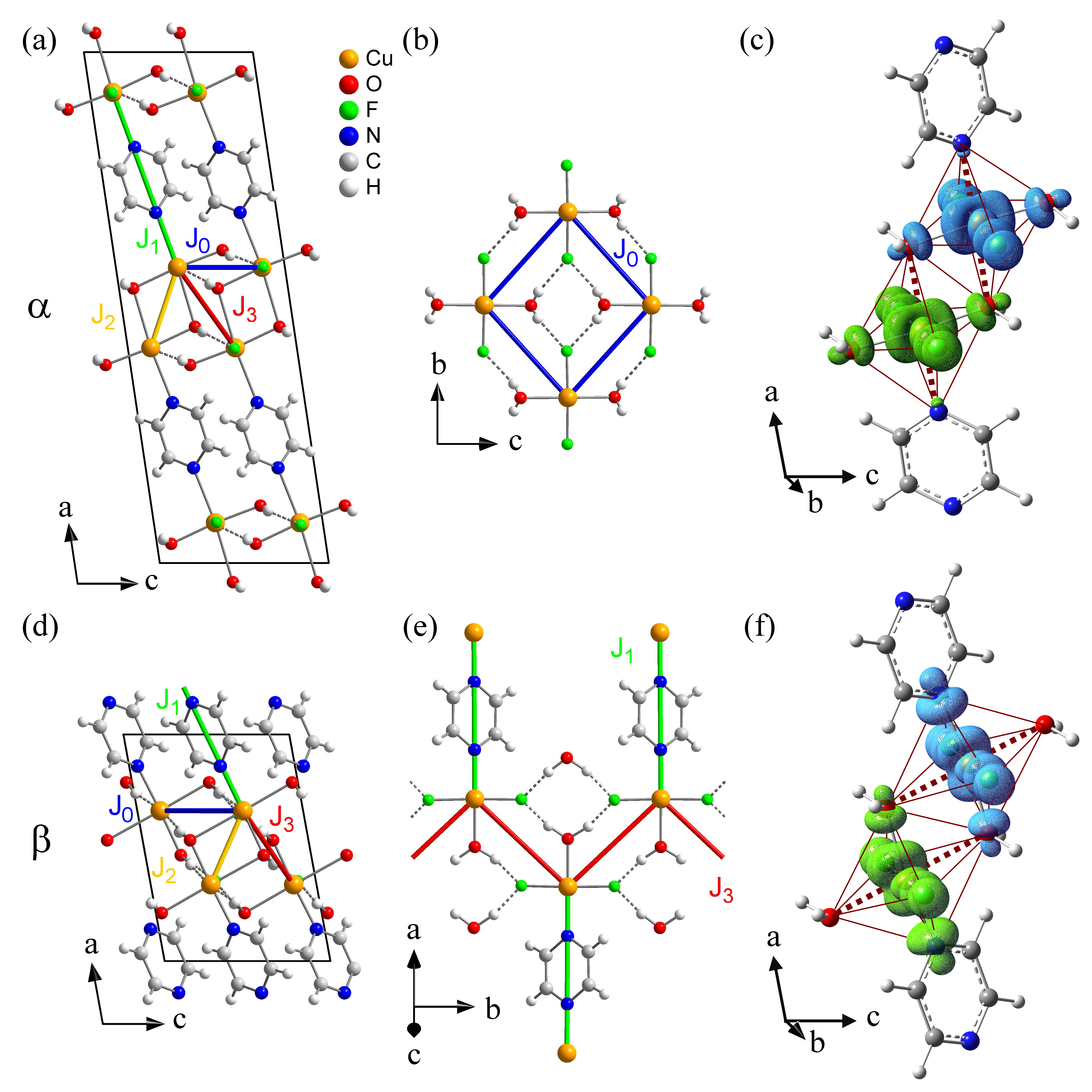}
\caption{Crystallographic structure of \bilayer in the $\alpha$-phase, showing views onto
(a) the $ac$ and (b) the $bc$ plane. The dominant exchange parameter, $J_0$, is 
mediated by Cu-O-H$\cdots$F-Cu superexchange paths. (c) Calculated spin-density distribution 
of the ground state, with spins up and down represented respectively in cyan and green. Dark red lines 
mark the coordination octahedron of the Cu$^{2+}$ ions and thick dashed lines 
the pseudo-Jahn-Teller axes. Structure in the $\beta$-phase showing views onto (d) the 
$ac$ and (e) the $ab$ plane. The dominant parameter, $J_3$, is mediated
by Cu-O-H$\cdots$F-Cu paths. (f) Both the pseudo-Jahn-Teller axes and the magnetic orbitals are 
reoriented at the phase transition.}
\label{fig:structure}
\end{figure}

At low pressures, \bilayer is a prototypical spin-1/2 square-lattice 
antiferromagnet with dominant in-plane magnetic exchange, $J_0$, and weak 
interlayer interactions \cite{lanza_cc_2014}. We demonstrate that \bilayer has 
three interlayer exchange parameters and present a full analysis of $J_1$, 
$J_2$, and $J_3$ in connection with Fig.~\ref{fig:structure}, but to complete 
the experimental analysis we take the result that $J_1$ and $J_2$ are relevant 
at low pressures. We have performed neutron diffraction measurements of the 
magnetic structure of \bilayer, detailed in Sec.~S3 of the SM \cite{sm}, which 
establish that $J_1$ is AFM and $J_2$ is FM. However, within the mean-field 
Random Phase Approximation (RPA) treatment \cite{scalapino_prb_1975} 
summarized in Sec.~S5 of the SM \cite{sm}, one may show that only the sum 
$|J_1| + |J_2| = 2J_{\perp}$ enters, and hence extract a single interlayer 
exchange parameter, $J_{\perp}$. 

For a full investigation of pressure dependence, we note that $g \mu_B B_c 
(P) = 4J_0 (P) + 2J_{\perp} (P)$ is the sum of all interaction strengths at a 
single Cu$^{2+}$ site, with $g = 2.42$ determined experimentally for $\bm B 
\parallel \bm a$ \cite{lanza_cc_2014}. $J_0 (P)$ and $T_N (P)$ can be used 
to determine one interlayer exchange parameter by employing the empirical 
relation 
\bea
\label{eq:QMC_2D_2}
J_{\perp} (P) = J_0 (P) \, e^{b - 4 \pi \rho_s / T_N (P)},
\eea
developed from quantum Monte Carlo (QMC) simulations for the spin-1/2 Q2D AFM 
Heisenberg model \cite{yasuda_prl_2005}, where $b = 2.43$ is a non-universal 
constant and $\rho_s = 0.183 J_0$ is the spin stiffness. This equation is valid 
for $0.001 \leq J_{\perp} / J_0 \leq 1$ and is obtained from a modified RPA 
(Sec.~S5 of the SM \cite{sm}).

With these equations as constraints, we obtain self-consistent 
values for $J_0(P)$ and $J_{\perp}(P)$ by computing the magnetic 
susceptibility. We perform QMC simulations using the ALPS open-source 
code \cite{bauer_jsm_2011}, as detailed in Sec.~S6 of the SM. The results 
of Fig.~\ref{fig:bilayer_data}(b) can be reproduced with 
quantitative accuracy at all fields and pressures by using a 
nearest-neighbor XXZ Hamiltonian on a simple cubic lattice, as 
illustrated in Fig.~\ref{fig:bilayer_data}(d) for the data at 
$P = 1.5$\,kbar. The spin-flop transition means that the SU(2) spin 
symmetry is broken down to U(1), and the measured $B_{\rm sf}$ value 
is obtained by setting $\Delta J^z_0 = J_0^z - J_0 = 0.09$\,K, i.e.~with 
a 1\% easy-axis anisotropy in $J_0$.

We show our results for $B_c (P)$ in Fig.~\ref{fig:bilayer_exchange}(b) 
and for $J_0(P)$ and $J_{\perp}(P)$ in Fig.~\ref{fig:bilayer_exchange}(d). 
Linear fits for the low-pressure ($\alpha$) phase yield $J_0(P) = 11.4(1)$\,K
 $-$ 0.34(1)\,$P$\,K/kbar and $J_{\perp}(P) = 0.33(1)$\,K $-$ 0.005(1)\,$P$\,K/kbar. 
Such a large coefficient for $J_0$ is quite extraordinary.
In Fig.~\ref{fig:bilayer_exchange}(c) we show the ratio $T_N/B_c$ 
as a function of pressure. Mean-field arguments predict both $T_N$ and $B_c$ 
to be proportional to the sum of all interactions and hence their ratio to 
be constant. However, quantum fluctuations in low-dimensional systems 
suppress $T_N$ (to zero in the 1D and 2D limits) but not $B_c$. Because 
$T_N/B_c$ is maximal for an isotropic (3D) system, our results imply that 
the Q2D system becomes slightly more 3D (i.e.~$J_{\perp}/J_0$ increases) 
with increasing pressure up to 18\,kbar. 

The discontinuous change at 18\,kbar marks a transition to a 
different low-dimensional magnetic phase. We find (below) that it is 
caused by a structural phase transition to a high-pressure $\beta$-phase. 
Here, the $J_3$ exchange becomes dominant, defining a system of AFM spin-1/2 
chains, while $J_{\perp}$ corresponds to the arithmetic mean of $J_0$, $J_1$, 
and $J_2$. For this Q1D case one has $g \mu_b B_c = 2 J_3 + 4 J_{\perp}$ and 
\bea
\label{eq:QMC_1D_2}
J_{\perp} = T_N/[4 c \sqrt{\ln(l J_3/T_N) + 0.5 \ln(\ln(l J_3/T_N))}],
\eea
where $c =$ 0.233 and $l =$ 2.6 \cite{yasuda_prl_2005}. A linear fit to the 
results for the $\beta$-phase yields $J_3 (P) = 12.7(1)$\,K 
$-$ 0.15(1)\,$P$\,K/kbar and $J_{\perp} (P) = 1.6(5)$\,K $-$ 0.03(1)\,$P$\,K/kbar; the 
coefficient of $J_3(P)$ is again anomalously large. 

To understand the giant pressure dependence of magnetic exchange in \bilayer, 
we have performed structural investigations by x-ray diffraction in order to 
benchmark first-principles calculations using spin-polarized DFT. As detailed 
in Sec.~S4 of the SM \cite{sm}, we made high-pressure single-crystal x-ray 
diffraction measurements at ambient temperature and powder measurements at 
5\,K. The unit-cell parameters and bond distances for different pressures are reported in 
Tables S1 and S2 of the SM \cite{sm} and full structural details are 
provided as crystallographic information files (CIFs). As represented in 
Fig.~\ref{fig:structure}, Cu$^{2+}$ ions are linked by OH$\cdots$F 
hydrogen bonds to form distorted square-lattice layers in the $bc$ plane. H$_2$O ligands further connect 
these into a bilayer and pyz molecules link the bilayers into a 3D 
coordination network. In the $\alpha$-phase 
[Figs.~\ref{fig:structure}(a)-\ref{fig:structure}(c)], the asymmetry in axial Cu coordination 
between the intrabilayer Cu$^{2+}$-H$_2$O bond and the interbilayer Cu-pyrazine direction 
is due to the "pseudo-Jahn-Teller" distortion. Upon compression, both axial ligands progressively approach Cu: Cu-N decreases from 2.40 to 2.30 \,\AA, whereas Cu-O decreases from 2.52 to 2.47\,\AA~(Table S2). Due to the stronger field of the pyz ligand, the Cu-N shortening is expected to affect more the metal stereochemistry. As shown below, this 
decrease is responsible for the giant pressure dependence of $J_0$.

\begin{figure}[t]
\includegraphics[width=7.5cm]{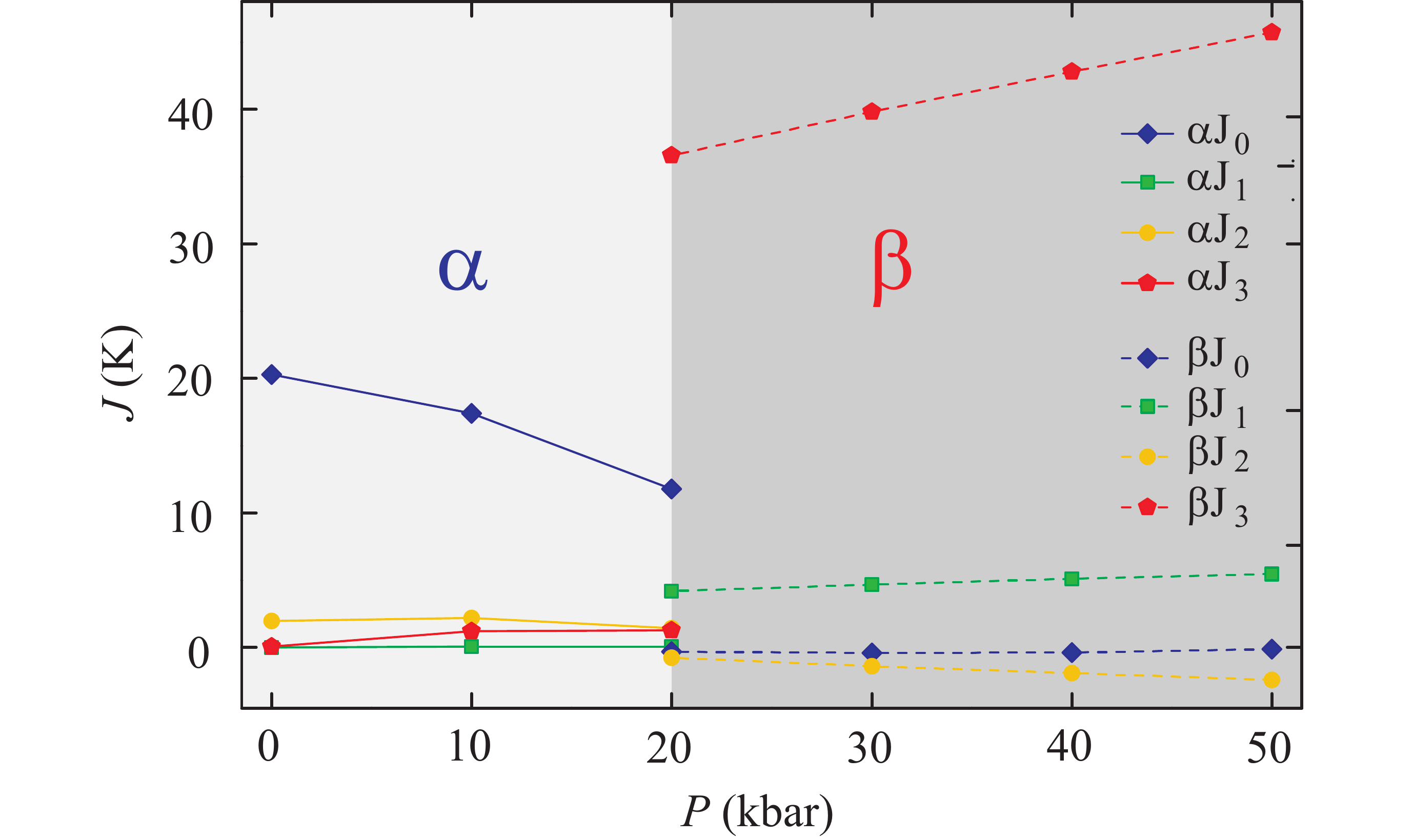}
\caption{Exchange parameters calculated as function of pressure for the 
$\alpha$- and $\beta$-phases using spin-polarized DFT.}
\label{fig:graphJvsP}
\end{figure}

A structural phase transition was observed at 18\,kbar. The high-pressure 
$\beta$-phase, shown in Figs.~\ref{fig:structure}(d)-\ref{fig:structure}(f), 
is characterized by a dramatic reduction of the Cu-N formally axial to 
2.1\,\AA~and an even stronger increase of the formerly equatorial Cu-O distances by up to 25\% (Table S2). This structural rearrangement indicates a change 
of the pseudo-Jahn-Teller axes [Fig.~\ref{fig:structure}(f)]. 
However, we note that the Cu-N distance remains longer than for a regular 
pyrazine coordination (2.05\,\AA). 

We use the lattice symmetry and approximate atomic positions at ambient 
pressure as input for geometry optimizations within periodic DFT calculations, 
which we perform using CRYSTAL14 \cite{Dovesi2014} as outlined in Sec.~S7 of 
the SM \cite{sm}. These reproduce all of the observed structural features, 
including their evolution as a function of pressures. They demonstrate that the $\beta$-phase is more stable than $\alpha$ for pressures above 18\,kbar, i.e.~the DFT 
calculations provide quantitative agreement on the critical pressure for 
the structural transition.

To investigate magnetic exchange in \bilayer, we identify the four Cu-Cu 
pathways shown in Fig.~\ref{fig:structure}. We obtain the exchange parameters 
from the energy differences between high- and low-spin states of dinuclear 
fragments, calculated using the GAUSSIAN09 package \cite{Frisch2009} with the 
procedure described in Ref.~\cite{Santos2016} and summarized in Sec.~S7 of 
the SM \cite{sm}. We find that the Cu$^{2+}$ ions have the highest spin 
densities, with the remaining fraction delocalized on the ligands. In the 
$\alpha$-phase, the magnetic orbitals involve F$^-$ and H$_2$O ligands
[Fig.~\ref{fig:structure}(c)] and the primary contribution to $J_0$ is from 
superexchange via Cu-O-H$\cdots$F-Cu paths, making Q2D magnetic layers 
that match the distorted structural square lattice (Fig.~\ref{fig:structure}(b) and 
Ref.~\cite{manson_cm_2008}). The other exchange paths, marked 
$J_1$, $J_2$, and $J_3$ in Fig.~\ref{fig:structure}(a), are poorly directed 
relative to the magnetic orbital and are small. 

The calculated magnetic exchange parameters are shown in 
Fig.~\ref{fig:graphJvsP}. DFT calculations without explicit account of 
correlation effects cannot in general obtain exchange parameters with 
quantitative accuracy, but their qualitative features contain essential 
physical insight. Most importantly, the giant decrease of $J_0$ in the 
$\alpha$-phase is in good qualitative agreement with experiment 
[Fig.~\ref{fig:bilayer_exchange}(d)]. Its microscopic 
origin lies mainly in the decrease of the axial Cu-N distance, which causes a 
systematic redistribution of the equatorial spin density of the magnetic orbital 
[Fig.~\ref{fig:structure}(c)] up to an orbital re-ordering and the occurrence of $\beta$-phase.
DFT indicates further that all of the subdominant exchange parameters are small. 
Although this places them below the resolution limits of our calculations \cite{thomas_exchange_2017}, it also supports the experimental analysis above. 
We draw attention to the trend visible in DFT that compression of the axial bonds enhances $J_3$ strongly, from 60\,mK at ambient pressure to 1.3\,K at 20\,kbar, without affecting 
$J_1$ or $J_2$ significantly. 

In the $\beta$-phase, the magnetic orbital revealed by the DFT spin density 
encompasses the two F$^-$ ions and the formerly axial water and pyz 
ligands [Fig.~\ref{fig:structure}(f)]. This orbital reorientation corresponds 
to the switch of the pseudo-Jahn-Teller axes and is responsible for the massive jumps in 
all of the exchange parameters (Fig.~\ref{fig:graphJvsP}). $J_3$ becomes the 
dominant exchange parameter [Fig.~\ref{fig:structure}(e)], whereas $J_1$ is 
significantly smaller (by a factor of 8 in our calculations). 
$J_0$ and $J_2$ are even weaker, 
because they involve water ligands lying normal to the magnetic orbital 
[Fig.~\ref{fig:structure}(f)]. Hence the system becomes Q1D due to the dominance of $J_3$. This 
pressure-induced switching of orbital orientation and system dimensionality 
is analogous to the transitions reported for the "monolayer" material 
\singlelayer \cite{halder_ac_2011,prescimone_ac_2012} and occurs despite the 
differences in Cu$^{2+}$ coordination. In neither case does the 
reorientation affect the covalently bonded part of the structure, although 
it modifies slightly the non-covalent interactions. We comment that, in 
contrast to the bilayer system, the monolayer one shows no significant 
pressure effect on the magnetic exchange away from the transition 
\cite{ghannadzadeh_prb_2013}.

Our combined experimental and theoretical results both demonstrate 
unequivocally and explain qualitatively the dramatic changes in the 
magnetic properties in \bilayer under applied hydrostatic pressure. There 
are two quite different types of change, namely (i) a giant but continuous 
decrease of the magnetic exchange parameter within the square lattice as the 
pressure is increased up to 18\,kbar and (ii) a discontinuous switching of 
the dimensionality of magnetic exchange from Q2D to Q1D above 18\,kbar. 

To explain results (i) and (ii), we have performed spin-polarized DFT calculations. Our magnetic 
calculations show that the key structural feature in the $\alpha$-phase
is the compression of the Jahn-Teller axes, which causes a progressive 
redistribution in the spin density of the magnetic orbital and thus 
the systematic and extremely strong reduction of the in-plane exchange. 
In the $\beta$-phase we find an abrupt switch in orientation of the magnetic 
orbital, causing the exchange to become dominated by the intrabilayer exchange $J_3$ and thus making a magnetic network that is Q1D, explaining the especially 
low $T_N/B_c$ in Fig.~\ref{fig:bilayer_exchange}(c). 

While our first-principles structural calculations for \bilayer under pressure 
are reliable at a quantitative level, our spin-dependent energetic calculations 
are not. Nevertheless, they do reproduce correctly the order of importance and 
the ratios of the exchange parameters at all pressures on both sides of the 
transition [Figs.~\ref{fig:graphJvsP} and \ref{fig:bilayer_exchange}(d)]. 
One key qualitative point is the DFT insight into the exchange parameters
$J_1$, $J_2$, and $J_3$, and specifically the fact that all of the subdominant 
parameters are weak, which allows us to disentangle them from a formalism 
based only on parameters $J_0$ and $J_{\perp}$. However, DFT does predict an 
increase of $J_3$ with pressure in the $\beta$-phase, in contrast to 
the decrease observed in experiment. Finally, a particularly valuable feature 
of our DFT results is to show the contributions of the different ligands 
involved in the superexchange paths, which is of vital importance in 
designing quantum magnets using metal-organic coordination polymers. 

At a fundamental level, our experiments provide extreme sensitivity for 
investigating questions such as the evolution of entanglement in the many-body 
wavefunction, in particular close to quantum phase transitions. Neutron 
spectroscopy allows a direct probe of magnetic correlations and excitations. 
Recent measurements on the monolayer material \singlelayer \cite{skoulatos_prb_2017} revealed that 
the orbital reorientation induces a first-order spin-wave to spinon transition 
of the magnetic excitations. Our findings show that \bilayer is a further 
excellent candidate for these studies, not only because the key physics occurs 
at accessible pressure, field, and temperature conditions, or even because of 
the dimensionality switching, but because of the enormous range of parameter 
ratios spanned continuously by this material.

At a more applied level, our measurements make \bilayer an 
important model system for benchmarking any theoretical approach aiming 
to provide a quantitative description of magnetic properties from first 
principles. Our results afford direct insight into the toolkit of 
metal-organic chemistry, in terms of the ligands and linking units giving 
maximal flexibility and control of magnetic exchange. Thus they provide 
an important step towards designing quantum magnets for applications in 
AFM spintronics, where we anticipate that pressure effects will be created 
using multiferroic substrate materials. For such devices to be realized in 
layered heterostructures, it is critical that the dimensionality switching 
should leave the effective low-dimensional magnetic system in the plane of 
the layer, which is the case in \bilayer but not in the monolayer material 
\singlelayer. 

In summary, we have observed and explained a giant pressure dependence 
of the magnetic exchange in the metal-organic quantum magnet \bilayer.
The combination of modern synthetic chemistry, high-precision physical 
measurements under extreme conditions, and state-of-the-art first-principles 
calculations allows a vital benchmarking of theoretical methods and provides 
a promising strategy for designing quantum materials with outstanding 
properties on demand. 

{\textit {Acknowledgments.}} We thank R. Schwartz for professional engineering 
of the pressure cells and fixtures used for this study and acknowledge fruitful 
discussions with T. Giamarchi and N. Qureshi. Computations were performed at 
the Universities of Geneva and Bern on the Baobab and UBELIX clusters and 
using the resources of the Theory of Quantum Matter Group in Geneva. This 
research was supported by the EU FP7/2007-2013 under Grant No.~290605, the 
European Research Council (ERC) under the EU Horizon 2020 research and 
innovation programme Grant No.~681654, and the Swiss National Science 
Foundation (SNSF) under Grants No.~200020\_150257 and 200020\_162861, as 
well as through the SINERGIA Network "Mott Physics Beyond the Heisenberg 
Model". J.G. was supported by an SNSF Early Postdoc Mobility fellowship 
during part of this work and M.S. by TRR80 of the German Physical Society 
(DPG). We further acknowledge funding for measurements performed at the 
NHMFL by Grant No.~DOE NNSA DE-NA0001979 with support by NSF Cooperative 
Agreement No.~DMR-1157490 and the State of Florida. 

\bibliographystyle{apsrev4-1_BW}
\bibliography{bilayer_refs}

\end{document}


\title{Supplemental Material for ``Giant pressure dependence and dimensionality 
switching in a metal-organic quantum antiferromagnet''}

\author{B. Wehinger, C. Fiolka, A. Lanza, R. Scatena, M. Kubus, A. Grockowiak,
W. A. Coniglio, D. Graf, M. Skoulatos, J.-H. Chen, J. Gukelberger, N. Casati,
O. Zaharko, P. Macchi, K. W. Kr\"amer, S. Tozer, C. Mudry, B. Normand, and Ch. 
R\"uegg}

\date{\today}

\maketitle

\setcounter{section}{0}
\setcounter{equation}{0}
\setcounter{figure}{0}
\setcounter{table}{0}
\setcounter{page}{1}
\renewcommand{\thesection}{S\arabic{section}}
\renewcommand{\theequation}{S\arabic{equation}}
\renewcommand{\thefigure}{S\arabic{figure}}
\renewcommand{\thetable}{S\arabic{table}}
\renewcommand{\bibnumfmt}[1]{[S#1]}
\renewcommand{\citenumfont}[1]{S#1}

\section{Sample}
\label{sec:sample}

Deuterated single crystals of \bilayer were grown by a method similar to that 
described in Ref.~\cite{lanza_cc_2014}. Deuterated ammonium fluoride (7.122\,g, 
192\,mmol) and D$_4$-pyrazine (CIL, 99.9\,\% D, 7.689\,g, 91\,mmol) were 
dissolved in 96\,ml water. A filtered solution of copper(II) nitrate made 
from copper(II) chloride (12.907\,g, 96\,mmol) and silver(I) nitride 
(32.616\,g, 192\,mmol) in 96\,ml water was added at $5^{\circ}$\,C. Upon slow 
evaporation, first \singlelayer crystals formed and then after several months 
\bilayer grew epitactically on these. Single crystals of \bilayer were 
separated and characterized by powder x-ray diffraction to verify their 
phase purity. The samples were stored in the mother liquor.

\section{Tunnel diode oscillator measurements}
\label{sec:tdo}

The TDO susceptometer consists of an oscillator circuit with a tunnel diode 
and a resonant $LC$ circuit where the sample is contained inside the inductor. 
Neglecting small parasitic components and lengthscales, the oscillation 
frequency is given by $\omega \propto 1/\sqrt{LC}$, where $L$ is modified by 
the susceptibility of the sample \cite{vandegrift1975,coniglio_arxiv_2010,
ghannadzadeh_rsi_2011}. Measurements using the piston-cylinder pressure cell 
were performed at the US National High Magnetic Field Laboratory's (NHMFL's) 35\,T, 32\,mm-bore resistive magnet (cell 12). 
The inductor was a single-layer coil, 800\,$\mu$m in diameter, of 25 turns 
wound with 28\,$\mu$m Cu wire. The piston cell was cooled by a $^4$He cryostat 
with a variable-temperature insert, where temperature is controlled by the 
flow of He gas entering the sample area from a valve at the base of the 
instrument. Susceptibility measurements were made at a temperature of 1.5\,K 
and pressures of 0.5, 1.5, 6.0, 13.6, and 17.9\,kbar. Our experiments at higher 
pressures were performed using a Moissanite anvil cell with 800\,$\mu$m culets; 
Ref.~\cite{graf_hpr_2011} describes a similar design. Here, the resonator coil 
had an outer diameter of 150\,$\mu$m, with 3 turns of 14\,$\mu$m-diameter 
copper wire (including insulation). Measurements were made in the NHMFL's 
superconducting magnet system SCM 2, where the base temperature is reached 
by condensing $^3$He gas to liquid using a 1\,K pot. These measurements 
were performed at a temperature of 0.4\,K and pressures of 16.1, 25.8, 31.9, 
and 37.1\,kbar. Daphne 7474 was used as pressure-transmitting medium 
\cite{murata_rsi_2008} and the pressure was determined \textit{in situ} 
from the fluorescence of a ruby crystal. 

\section{Neutron diffraction}
\label{sec:neutron}

Neutron diffraction measurements were performed on deuterated single crystals 
of \bilayer using the TriCS single-crystal diffractometer at the Swiss 
Spallation Neutron Source, SINQ, at the Paul Scherrer Institute. Datasets 
were collected at ambient pressure at temperatures of 7\,K and 1.5\,K, which 
lie above and below the magnetic transition. The wavelengths used were 
1.18\,\AA~and 2.32\,\AA, respectively from Ge(311) and PG(002) monochromators.
The magnetic propagation vector of \bilayer is $k = (0\,0\,0)$, i.e.~the 
magnetic peaks coincide with the structural Bragg peaks. The system is a 
N\'eel antiferromagnet within the $bc$ plane (Fig.~3(b) of the main text). 
We find that the two planes within a bilayer unit are ferromagnetically 
coupled, meaning that $J_2 < 0$. By contrast, the bilayer units are 
antiferromagnetically coupled ($J_1 > 0$). The system has a fully ordered 
moment of 1\,$\mu_B$, which we established by normalizing our results at 
1.5\,K with the structural data collected at 7\,K.

\section{X-ray diffraction}
\label{sec:xray}

\begin{ruledtabular}
\begin{table*}[tp]
\caption{Structural parameters of \bilayer obtained by x-ray diffraction at 
different pressures and temperatures. Values reported at 5\,K were obtained 
from high-resolution powder diffraction and confirm the absence of a 
low-temperature phase. Values at room temperature were obtained from 
refinements of single-crystal x-ray diffraction data.}
  {
    \setlength\tabcolsep{1.5pt} 
    \begin{tabular}{llllllll}
     Pressure (kbar) & 0 & 0 & 10 & 18 & 33 \\
     Temperature (K) & 5 & 298 & 298 & 298 & 298  \\
     \hline     
     Space group & $I2$\slash$c$ & $I2$\slash$c$ & $I2$\slash$c$ & 
     $I2$\slash$c$ & \\
     $a$ (\AA) &  20.842(2) & 21.0200(3) & 20.76(2) & 20.753(16) &  \\
     $b$ (\AA) & 7.5562(7) & 7.5530(1) & 7.5022(8) & 7.5170(6) &  \\
     $c$ (\AA) & 6.8008(6) & 6.8810(1) & 6.6300(5) & 6.4475(4) & \\ 
     $\beta$ ($^{\circ}$) & 99.3171(7) & 98.456(2)  & 101.45(3) & 103.21(2) & \\
     \hline
     Space group & & & &  $P2_1$\slash$c $ & $P2_1$\slash$c $\\
     $a$ (\AA) & & & &  9.8(5) & 9.230(12) \\
     $b$ (\AA) & & & & 7.429(11) &  7.5743(8)  \\
     $c$ (\AA) & & & & 6.95(4) &  6.6446(10) \\
     $\beta$ ($^{\circ}$) & & & & 98(2)  &  100.32(5)  \\
    \end{tabular} 
  }
   \label{tab:diffraction}
\end{table*}
\end{ruledtabular}

\begin{ruledtabular}
\begin{table*}[tp]
\caption{Bond distances (\AA) in \bilayer at room temperature (298\,K) 
obtained from refinements of single-crystal x-ray diffraction data.
The designations equatorial (eq.) and axial (ax.) are relative to the 
pseudo-Jahn-Teller axis (Figs.~3(c) and 3(f) of the main text).}
  {
    \setlength\tabcolsep{1.5pt} 
    \begin{tabular}{llllllll}
     & $\alpha$-phase & & & & $\beta$-phase & & \\
     Pressure (kbar) & & 0 & 10 & 18 & & 18 & 33 \\
     \hline
    Cu - F	    & eq. & 1.899(1) & 1.888(4) & 1.902(4) & eq. & 1.863(6) & 1.894(6) \\
    		    & eq. & 1.893(1) & 1.891(4) & 1.895(4) & eq. & 1.871(5) & 1.906(5) \\
    Cu - O	    & eq. & 1.966(1) & 1.960(5) & 1.965(5) & ax. & 2.29(4)  & 2.27(1)  \\
    (b-c plane)     & eq. & 1.993(1) & 1.967(6) & 1.983(6) & ax. & 2.49(5)  & 2.43(1) \\
    Cu - N	    & ax. & 2.404(2) & 2.38(2)  & 2.30(1)  & eq. & 2.1(1)   & 2.01(2)  \\
    Cu - O	    & ax. & 2.522(1) & 2.51(1)  & 2.47(1)  & eq  & 2.1(1)   & 1.96(3)  \\
    \end{tabular} 
  }
   \label{tab:ligand-dist}
\end{table*}
\end{ruledtabular}

A single-crystal sample of \bilayer was compressed in a Merrill-Bassett 
diamond-anvil cell equipped with 0.5\,mm diamonds and a steel gasket, 
pre-indented to 0.080\,mm and with a 0.20\,mm hole diameter. Daphne oil 
was used as the pressure-transmitting medium and the pressure calibrated 
by the ruby-fluorescence method. The diffracted intensities were collected 
at 10, 18, and 33\,kbar at room temperature using a Rigaku Oxford Diffraction 
SuperNova area-detector diffractometer, where the Mo K$\alpha$ radiation 
($\lambda$ = 0.71073\,\AA) is monochromated by mirror optics. For each 
pressure, in total 369 frames were collected in 1$^{\circ}$ steps, with 
respective exposure times of 20 or 60 seconds for low or high scattering 
angles, and a fixed sample-detector distance of 68.0\,mm; CrysAlisPro 
\cite{Crysalis2017} was used for the data collection, data reduction, and 
empirical absorption-correction. The structural model was refined by the 
least-squares method using SHELXL-2016 \cite{Sheldrick2014}. D atoms on the 
pyrazine ring were assigned geometrically and refined using a riding model 
with an isotropic thermal parameter equal to 1.2 times that of the 
corresponding parent atom. D atoms of the coordinating water molecules were 
refined with a restrained O-D distance for the $\alpha$-phase and fixed at 
their geometrically optimized coordinates for the $\beta$-phase. 

Powder x-ray diffraction at ambient pressure and a temperature of 5\,K 
was conducted at the Materials Science beamline at the Swiss Light Source 
\cite{Willmott2013}. The sample was loaded in a 0.3\,mm-diameter glass 
capillary, aligned to ensure rotation on axis, and fixed in a modified 
Janis cryostat. The temperature was driven by blowing He on the sample and 
controlled by the local infrastructure, which included a thermocouple at the 
exit of the He flow. The wavelength was calibrated as 0.56332\,\AA, using 
the Si standard 640d from NIST, and instrument function was also calibrated 
against the diffraction of a similarly-sized capillary containing powders 
of Na$_2$Al$_2$Ca$_3$F$_{14}$. A Mythen microstrip detector was used for the 
measurements. Data were collected at several temperatures while heating 
the sample back from 5\,K, which showed evidence for the presence of 
\singlelayer \cite{manson_cm_2008}, plus some small spurious peaks from the 
apparatus that were easy to identify because they do not move with temperature. 
The $\alpha$-phase was observed as the primary constituent, with no evidence 
for a structural transition in the temperature range analyzed.

Full results for the structure of \bilayer are reported in Tables 
\ref{tab:diffraction} and \ref{tab:ligand-dist} and in the associated 
crystallographic information files. For some pressure points, only the 
unit cell was determined. In our measurements at 18\,kbar, close to the 
phase transition, we detected two diffraction patterns simultaneously from 
the sample in the DAC, one corresponding to the $\alpha$-phase and the other 
to the high-pressure $\beta$-phase. At 33\,kbar, only the $\beta$-phase 
remained visible. At higher pressures, the single crystal was damaged too 
severely for further investigation.

\section{Random-Phase Approximation for Quasi-2D Susceptibilities}
\label{sec:rpa}

To provide a qualitative explanation for the forms of Eqs.~(1) and (2) of 
the main text, and to simplify the parameter space of our QMC calculations,
we discuss briefly the RPA form of the susceptibility in a strongly spatially 
anisotropic spin system. For a system of very weakly coupled magnetic units, 
the effect on a single unit of all its neighbors may be modelled as an 
effective static field, which is the origin of the name RPA. This field, 
which is staggered according to the magnetic ordering pattern of the units, 
can be added to the uniform external magnetic field to give a single, 
effective field $h_i$ on each site. Considering for illustration the Q2D case 
(the $\alpha$-phase of \bilayer), we take the interlayer exchange parameter $J_1$ and 
$J_2$ to obey $J_0 \; \gg \; |J_{1}|,\,|J_{2}^{}| \; > 0$. Working in the 
classical mean-field approximation, at the level of linear response one 
may express the full 3D susceptibility as 
\bea
\label{eq:RPA3D}
\chi^{3D}_{\boldsymbol{q}} = \frac{\chi^{2D}_{\boldsymbol{q}_{\parallel}} \left[ 
1 - (J_{1} + J_{2}) \cos{q}_{\perp} \chi^{2D}_{\boldsymbol{q}_{\parallel}} \right]}
{1 - (J_{1}^2 + J_{2}^2 + 2 J_{1} J_{2} \cos{2q}_{\perp}) 
[\chi^{2D}_{\boldsymbol{q}_{\parallel}}]^{2}}\,,
\eea
where ${\bm q} \equiv ({\bm q_{\parallel}^{}}, q_{\perp}^{})$ is the wavevector, 
$\perp$ indicates the (stacking) direction perpendicular to the 2D layer 
and $\chi^{2D}_{\bm q_{\parallel}}$ is the 2D AFM susceptibility. On taking the 
uniform susceptibility, $q_{\perp}$ is determined by the signs of $J_1$ and 
$J_2$ and from Ref.~\cite{chakravarty_prb_1989} one has 
\bea\label{eq:2dchisig}
\chi^{2D}(T) = c_2^{}\frac{T}{J_0^2}\,e^{4\pi\rho_{s}/T},
\eea
where $c_2$ is a non-universal constant and $\rho_s$ the 2D spin stiffness.  
Solving the pole equation given by the denominator of $\chi^{3D}_{\boldsymbol{q}}$ 
yields
\bea\label{eq:2dTN}
T_{N} = \frac{4\pi\rho_{s}}{- \ln 2c_2 -\ln \frac{|J_1| + |J_2|}{2J_0} - 
\ln \frac{T_{N}}{J_0}},
\eea
or equivalently 
\bea
\label{eq:RPA}
(|J_1| + |J_2|)/2 = J_{\perp} = \frac{J_0^2}{2 c_2 T_N} e^{-4 \pi
 \rho_s / T_N}.
\eea
This exponential dependence of $J_\perp$ on $1/T_N$ is the basis of Eq.~(1) of 
the main text and the special case $|J_1|$ = $|J_2|$ agrees with the results 
of Ref.~\cite{yasuda_prl_2005}. We comment that the parameters we obtain for 
the $\alpha$-phase, $J_0/J_\perp \simeq 35$, mean that mean-field RPA, Eq.
(\ref{eq:RPA}), is already at the limit of its quantitative applicability, 
and hence we use the empirical formula of Ref.~\cite{yasuda_prl_2005} 
(Eq.~(1) of the main text). The analogous treatment of a Q1D system 
\cite{scalapino_prb_1975} is the basis of the modified RPA giving Eq.~(2) 
of the main text.

By the nature of their exchange paths, the interlayer exchange parameter $J_1$ and 
$J_2$ in the $\alpha$-phase of \bilayer have unrelated values. However, 
for the QMC analysis of our experimental data, discussed in Sec.~S5, we 
reduced the number of free parameters by the somewhat arbitrary assumption 
$|J_1| = |J_2|$. The RPA treatment demonstrates that the exact values of 
these weak exchange parameters have no significant influence on the qualitative 
aspects of our results, at least as long as neither vanishes. Although 
we have established the signs of $J_1$ and $J_2$ for \bilayer by our 
neutron diffration measurements of the magnetic structure (Sec.~S4), the 
RPA treatment demonstrates that Eqs.~(1) and (2) of the main text are valid 
for small FM, AFM, and alternating exchange along the stacking direction(s).

\section{Quantum Monte Carlo simulations}
\label{sec:qmc}

To model the susceptibility measured in the $\alpha$-phase, despite the unknown 
values of the weak interlayer exchange parameters $J_1$ and $J_2$, we invoke the RPA 
results of Sec.~S3 to justify proceeding by setting $J_1 = |J_2| = J_{\perp}$. 
Thus we perform QMC simulations of a spin-1/2 system on a primitive cubic 
lattice with Hamiltonian
\bea
\label{eq:qmchamiltonian}
\widehat{H} = && \sum_{\boldsymbol{n},m,a} \Big( \sum_{\mu = 1,2} J_0^a
\widehat{S}^{a}_{\boldsymbol{n},m} \widehat{S}^{a}_{\boldsymbol{n}+\boldsymbol{\delta}_{\mu},m}
+ J_{\perp} \widehat{S}^{a}_{\boldsymbol{n},m} \widehat{S}^{a}_{\boldsymbol{n},m + 1}
\nonumber \\ && \;\;\;\;\;\;\;\;\;\;\;\;\;\;\;\; - \; h^{a}_{\boldsymbol{n},m} 
\widehat{S}^{a}_{\boldsymbol{n},m} \Big),
\eea
where ${\bm n}$ is the 2D vector specifying the sites in one layer, 
$\mu$ labels the two orthogonal unit vectors spanning the square plane 
[$\boldsymbol{\delta}_{\mu} = (1,0), (0,1)$], and $m$ is the index along 
the stacking direction of the layers. The label $a = x$, $y$, $z$ enumerates 
the spin components, which we differentiate because the presence of the 
spin-flop transition indicates a non-Heisenberg anisotropy in the interaction.
We choose the $z$-direction to correspond to the crystallographic $a$-axis, 
which is aligned with the magnetic field, i.e.~$h^{z}_{\boldsymbol{n},m} = h$ and 
$h^{x}_{\boldsymbol{n},m} = h^{y}_{\boldsymbol{n},m} = 0$. 

We focus on the susceptibility data obtained at $P = 1.5$\,kbar and set the 
exchange parameters to the values deduced from the corresponding $B_c$ and 
$T_N$. Because the interaction anisotropy is expected to be weak, we neglect 
any anisotropy in the interlayer exchange parameters and take $J_\perp = 0.34$\,K. For 
the intralayer exchange parameters we take $J^x = J^y = 10.83$\,K. For the $g$-factor 
we use the value reported under ambient conditions, $g = 2.42$ 
\cite{lanza_cc_2014}. To reproduce the observed spin-flop field, $B_{\rm sf} 
\simeq$ 1.2\,T, we find that the easy-axis anisotropy in $J_0$ should satisfy 
$J_0^z - J_0^{x,y} = 0.09$\,K, coeresponding to an anisotropy of order 1\%. 

For each field value shown in Fig.~1(d) of the main text, we performed 
10$^4$ Monte Carlo sweeps for thermalization and 8$\times 10^5$ sweeps after 
thermalization. The spin-flop transition is sensitive to finite-size effects 
and a rather large number of spins is required for an accurate description. 
For this reason we computed the susceptibility, magnetization, Binder cumulant, 
energy, and their auto-correlation times for different system sizes at the 
transition and verified that the value of the transition field, $B_{\rm sf}$, 
had converged to within 0.05\,T as a function of system size, which was 
varied up to 64$\times$64$\times$64 spins. The stochastic error (one standard 
deviation) in the QMC susceptibility is 5\% at the spin-flop transition and 
less than 0.3\% elsewhere.

\section{Ab initio calculations}
\label{sec:abinitio}

Optimization of the structure at different values of the pressure was 
performed for both $\alpha$- and $\beta$-phases in their ferromagnetic 
configuration using periodic DFT, as implemented in the CRYSTAL14 package 
\cite{Dovesi2014}. The unrestricted functional B3LYP \cite{Becke1993,Lee1988} 
and the basis set 6-31G(d,p) \cite{Gatti1994} were used. 

To compute the magnetic exchange, closed-shell fragments of the structure 
were selected to separate each possible exchange path. No further geometry 
optimization was performed on these fragments, because small variation in the 
geometry induced, for instance, by the absence of packing forces, can cause 
relevant changes of the exchange parameter. The use of fragments allows 
one to employ the extended 6-311G(2d,2p) basis set. DFT calculations using 
GAUSSIAN09 software \cite{Frisch2009} were carried out in the gas phase to 
obtain the energies of high- and broken-symmetry low-spin states of the 
fragments. Each of the exchange parameter could then be deduced 
separately from the energy differences, whereas in calculations with periodic 
boundary conditions some of the pathways shown in Fig.~3 of the main text 
would be mutually entangled.

We chose the functional B3LYP because it includes the correct amount of 
Hartree-Fock exchange for a proper description of high- and low-spin states 
\cite{Ruiz1999}, and has been shown in previous studies to predict correctly  
the exchange parameters in Cu$^{2+}$-based metal-organic quantum 
magnets \cite{Santos2016,Brambleby2017}. For comparison, we also tested other 
compatible functionals, such as B3P86 and M06 \cite{Zhao2008}, and obtained 
analogous results, whereas functionals based purely on the generalized 
gradient approximation failed to produce meaningful exchange parameters. 

\bibliographystyle{apsrev4-1_BW}
\bibliography{bilayer_refs}